\documentclass{article}
\usepackage[utf8]{inputenc}
\usepackage{graphicx}
\usepackage{tabularx}
\def\be{\begin{equation}}
\def\ee{\end{equation}}
\def\bea{\begin{eqnarray}}
\def\eea{\end{eqnarray}}

\usepackage{url}

\usepackage{caption}
\usepackage{lineno}
\usepackage{float}

\title{Muons for cultural heritage}

\author{M. Moussawi$^{1,*}$ \and A. Giammanco$^{1}$ \and V. Kumar$^{1}$ \and M. Lagrange$^{1}$  }

\date{$^1$Centre for Cosmology, Particle Physics and Phenomenology (CP3), Universit\'e catholique de Louvain, Chemin du Cyclotron 2, B-1348 Louvain-la-Neuve, Belgium \\
     [3ex] \today}

\begin{document}

\maketitle

{\raggedleft CP3-23-50 \\
}

\begin{abstract}
Non-destructive subsurface imaging methods based on the absorption or scattering of photons or neutrons are becoming increasingly popular in cultural asset conservation. However, these techniques are limited by physical and practical issues: their penetration depth may be insufficient for large items, and they usually necessitate transferring the objects of interest to specialised laboratories. 
The latter issue is recently being addressed by the development of portable sources, but artificial radiation can be harmful and is thus subjected to strict regulation. 
Muons are elementary particles that are abundantly and freely created in the atmosphere by cosmic-ray interactions. Their absorption and scattering in matter are respectively dependent on the density and elemental composition of the substance they traverse, suggesting that they could be used for subsurface remote imaging. This novel technique, dubbed "muography," has been used in applications ranging from geophysics to archaeology, but has remained largely unexplored for a wide range of cultural heritage objects that are small by muography standards but whose size and density are too large for conventional imaging methods.
This document outlines the general arguments and some early simulation studies that aim at exploring the low-size limit of muography and its relevance for cultural heritage preservation.
\\
\\
{\it Proceedings of the Muon4Future Conference, 29-31 mai. 2023 at Venice, Italy. Submitted to Proceeding in Science.}
\end{abstract}

\section{Introduction}
\label{sec:intro}
\begin{figure}[H]
\centering
\begin{minipage}{13pc}
\centering
\includegraphics[width=12pc]{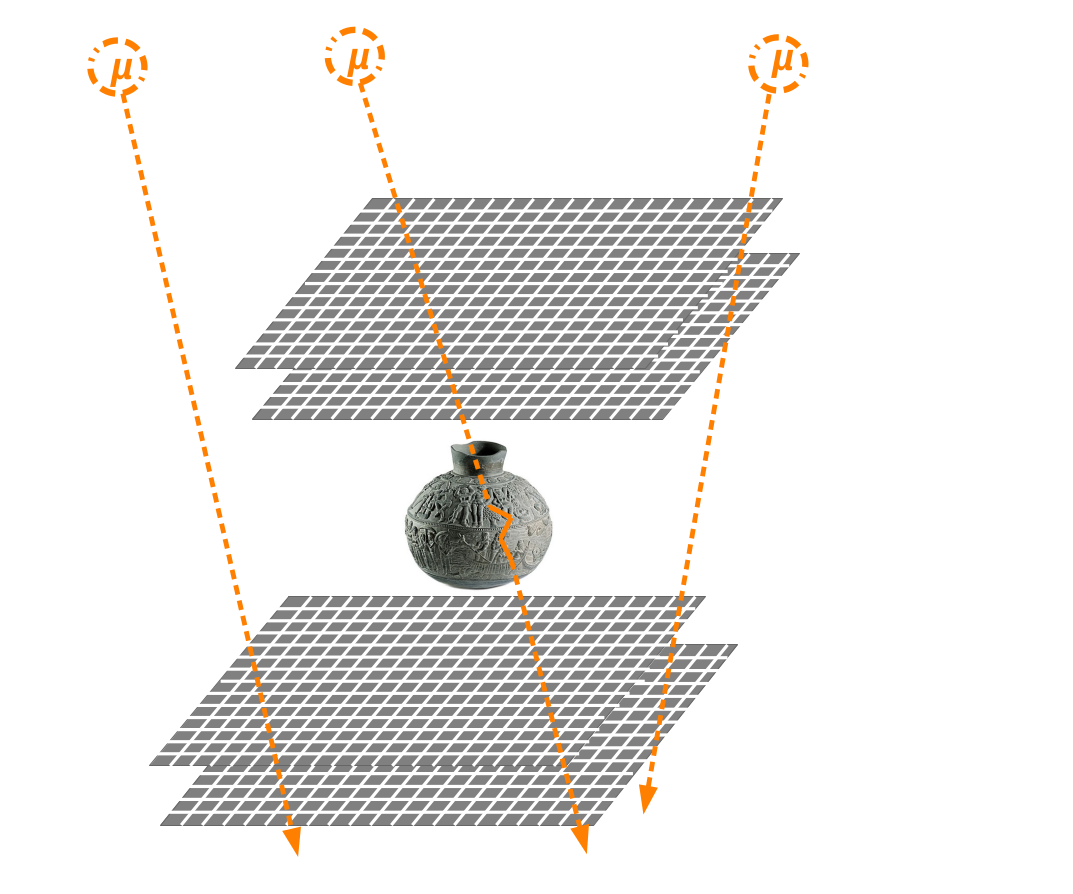}
\caption{In scattering muography, the object under investigation is ”sandwiched” between muon trackers.} 
\label{sketch_scatter}
\end{minipage}\hspace{2pc}
\begin{minipage}{13pc}
\centering
\includegraphics[width=12pc]{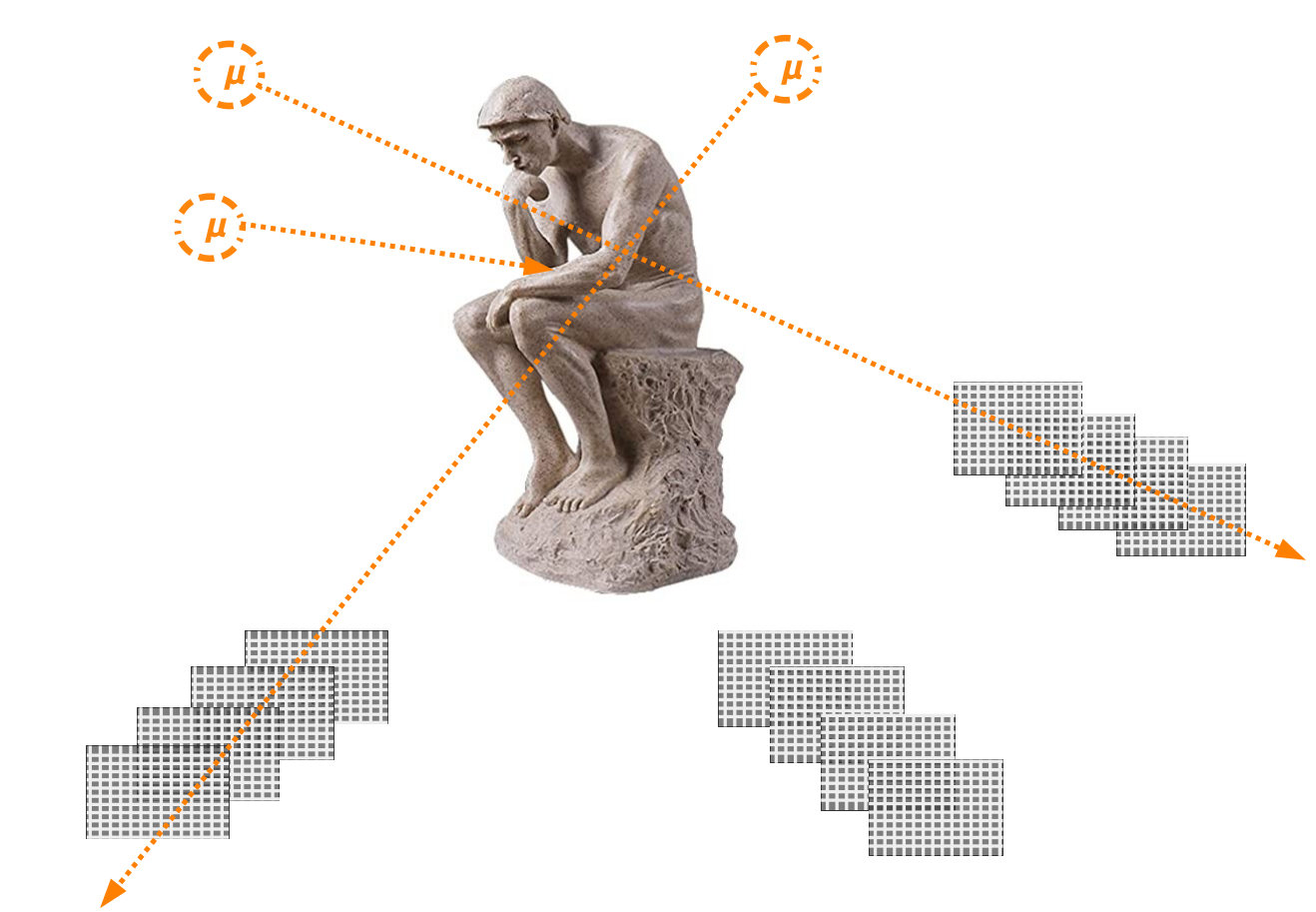}
\caption{In absorption muography, muon trackers are downstream of the object of interest; 3D imaging can be obtained by combining multiple viewpoints.}
\label{sketch_absorpt}
\end{minipage} 
 \end{figure}

Imaging methods based on X-rays have been widely used in the context of cultural heritage preservation~\cite{morigi2010application} due to their ability to penetrate various materials. However, X-ray imaging has limitations when dealing with large or dense objects like compact stone or metal, since they do not penetrate deep enough. Alternative radiation types, such as MeV-range X-rays and neutrons, offer some improvement but face challenges in transporting valuable objects to specialized imaging facilities due to size, weight, and preservation concerns.
Various portable setups, such as X-ray fluorescence analysis (XRF)~\cite{bezur2020handheld,ridolfi2012portable} and portable X-ray computed tomography (CT) systems~\cite{Albertin2019}, are available for cultural heritage studies, but they have limitations in depth penetration and radiation hazards. Neutron sources~\cite{kerr2022neutron} offer greater depth~\cite{kardjilov2018advances}, but raise concerns about material activation. A recent advancement using a portable proton accelerator~\cite{taccetti2023machina} shows promise but also suffers from radiation hazard concerns.
In contrast, muography~\cite{Bonechi:2019ckl}, which utilizes muons ($\mu$), elementary particles naturally generated by cosmic-ray interactions in the atmosphere, represents a promising solution. Cosmogenic muons have remarkable penetrating capabilities, making them ideal for sub-surface imaging in a variety of contexts including, as we argue in this paper, cultural heritage applications. This technique includes two main methods: scattering-based and absorption-based, sketched in Fig.\ref{sketch_scatter} and Fig.\ref{sketch_absorpt} respectively. Absorption-based muography measures muons absorption rate within materials, providing insights into their density and composition, while scattering-based muography utilizes the diffusion of muons to discriminate elements in multi-material objects~\cite{Borozdin2003}.

Muography has proven to be highly effective in investigating cultural heritage sites. 
The ScanPyramids project, utilizing absorption-based muography, made the headlines by revealing an unexpected low-density anomaly deeply inside Khufu's Great Pyramid in 2017~\cite{Morishima2017}, and then precisely characterizing a previously unknown corridor in 2023~\cite{procureur2023precise}, which has been confirmed by visual inspection via an endoscope. 
In another example, density anomalies (potentially posing safety hazards) have been found in a rampart of a defensive wall of Xi’an (China) in 2022~\cite{Xian-walls}.
Furthermore, scattering-based muography has been proposed to search for iron chains within the brickwork of the Florence cathedral's dome in Italy~\cite{Guardincerri2018}, and a proof-of-principle test on a mock-up wall was successfully conducted to demonstrate the conceptual validity of the method. 

While most examples so far are applications to very large volumes of interest, this paper advocates for the adoption of portable and safe muography as a promising imaging approach for cultural heritage studies in a regime that is new for muography (relatively low size) while being beyond reach for methods based on other radiation sources. A preliminary simulation study using Geant4~\cite{GEANT4} illustrates the potential applications and limitations of muography.

\begin{figure}[H]
\centering
\begin{minipage}{13pc}
\includegraphics[width=8pc]{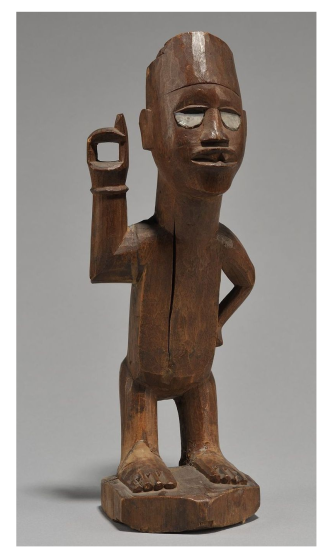}
\caption{Picture of the wooden statue at the Africa Museum of Tervuren. From project TOCOWO (\protect\url{https://tocowo.ugent.be/}).}
\label{statue}
\end{minipage}\hspace{2.5pc}
\begin{minipage}{13pc}
\includegraphics[width=6pc]{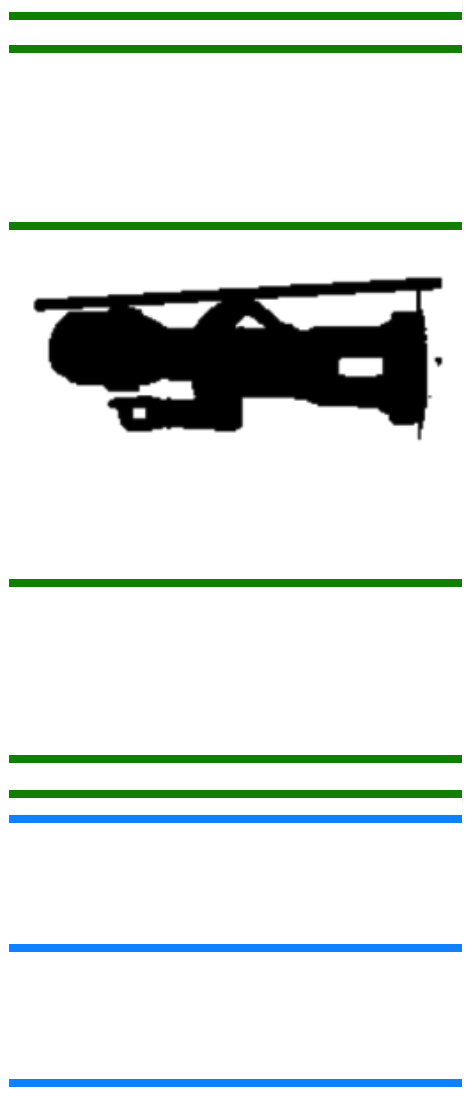}
\caption{Geant4 simulation setup; green and blue panels represent the scattering and absorption setups, respectively.}
\label{setup}
\end{minipage} 
\end{figure}

\section{Simulated case studies}
\label{sec:simulation_case}

Each of the two muography techniques has its own sensitivity, applicability, and limits.

In absorption muography, a single muon tracker is able to measure the 2D projection of matter density, and the combination of measurements from different viewpoints can give a 3D density map. 
However, it provides no material discrimination apart from density, and small-size or low-density objects do not stop enough muons to provide sufficient contrast. 

In scattering muography, at least two muon trackers are needed, upstream and downstream of the object of interest, to reconstruct the $\mu$ trajectory before and after passing through it. 
This method naturally yields 3D information, and is sensitive to elemental composition because the width of the scattering angle distribution is a function of atomic number Z. 
However, it is
impractical for human-sized statues, as the object of interest must fit between the two trackers.
Either the object of interest is moved inside the set-up, or a rather complex installation of the detectors must be performed around the object. 
Therefore, this method is appropriate for relatively low-size objects. 

We perform a Monte Carlo simulation using CRY~\cite{CRY} to generate muons and a Geant4~\cite{GEANT4} model of an African statue (Fig.~\ref{statue}) made of hardwood. This object, 40~cm tall, has been studied with X-rays but its size nears the limit of that method, while it is very small by muography standards.
To investigate the potential of muography for material identification, we scale the statue's size by factors two and four and introduce hidden cylinders of different materials within its internal structure, as summarized in Table~\ref{tab-Simulation}.
In this first exploratory study, we model an ideal detector (i.e. with 100\% efficiency and perfect resolution) made up of nine planes, as illustrated in Fig.~\ref{setup}. Among these planes, the six indicated in green surround the target and are used for scattering reconstruction, while the remaining three (in blue) are used in the absorption reconstruction study. 



\begin{table}[ht]
\centering
\begin{tabular}{|c|c|c|c|c|}
\hline
\bf Scenario & \bf Statue size [cm$^3$]  & \bf Cylinder material  & \bf Cylinder radius [cm] \\ 
\hline

I (a) & $80 \times 30 \times 30$ & $/$ & $/$    \\ \hline
I (b) & $80 \times 30 \times 30$ & Air & 5 \\ \hline
I (c) & $80 \times 30 \times 30$ & Bronze bar & 5  \\ \hline
II~~ & $160 \times 60 \times 60$ & Bronze bar & 10 \\ \hline
\end{tabular}
\caption{Different simulation scenarios.}
\label{tab-Simulation}
\end{table}



{\bf Scattering reconstruction}

\begin{figure}[H]
    \centering
    \includegraphics[width=0.32\textwidth]{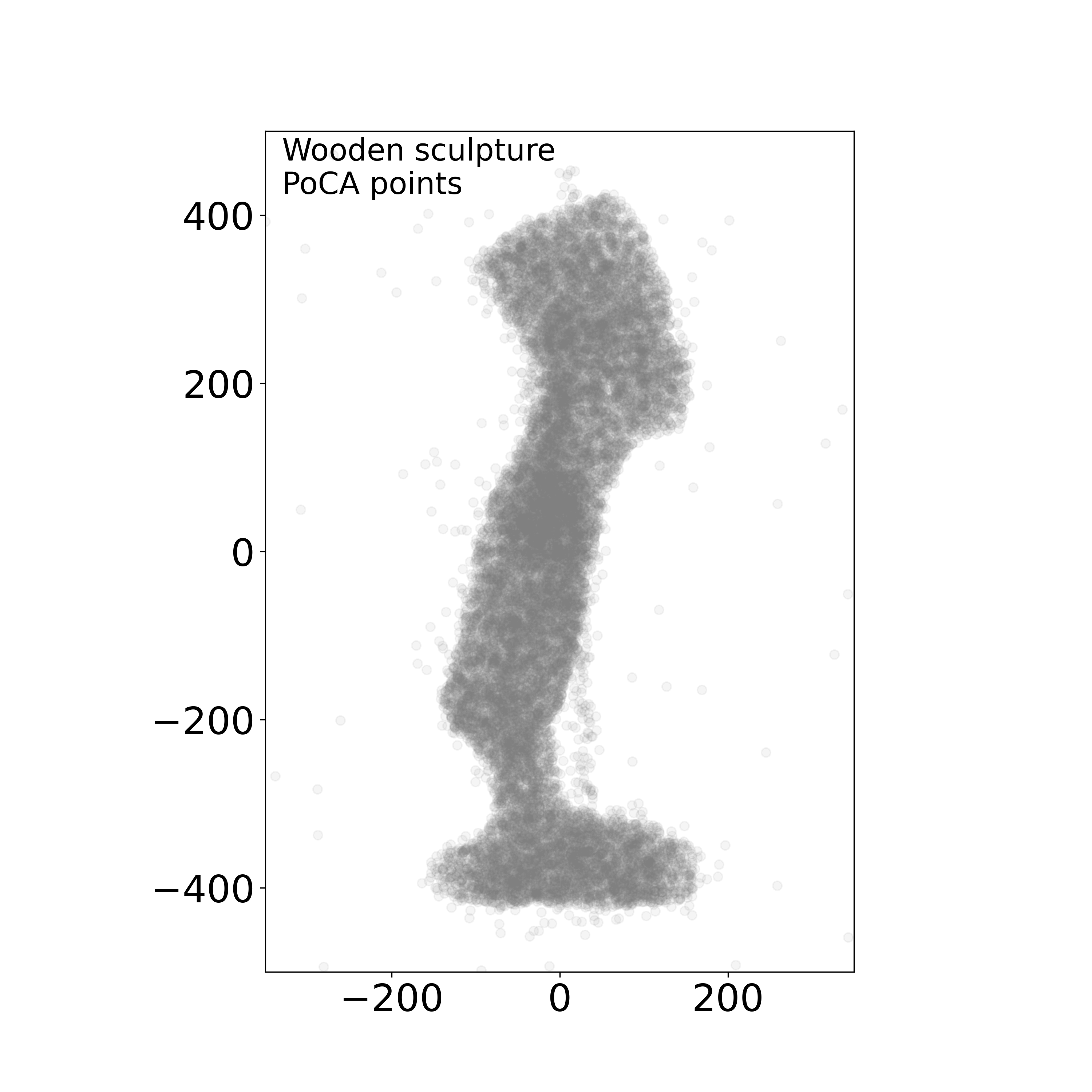}
    \includegraphics[width=0.32\textwidth]{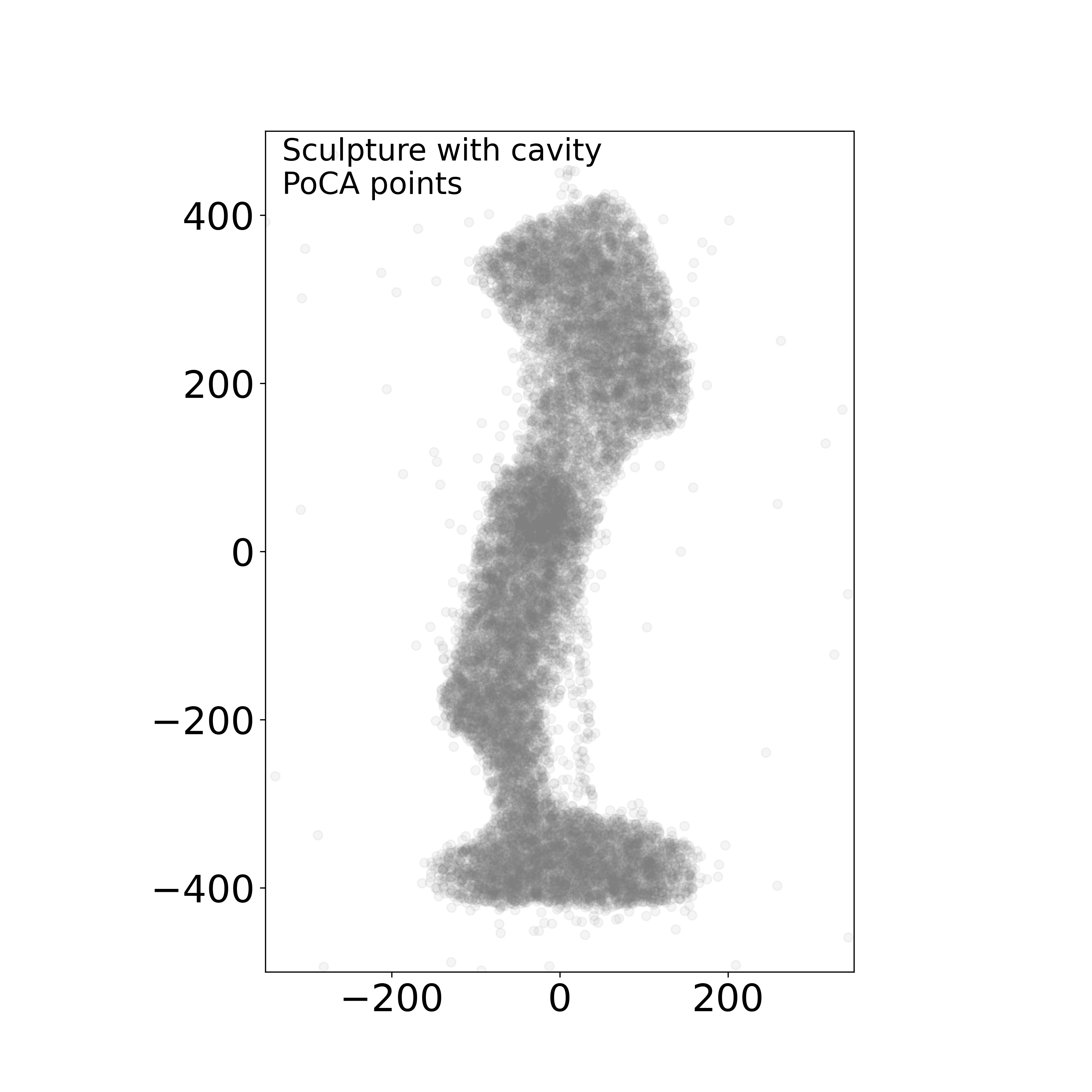}
    \includegraphics[width=0.32\textwidth]{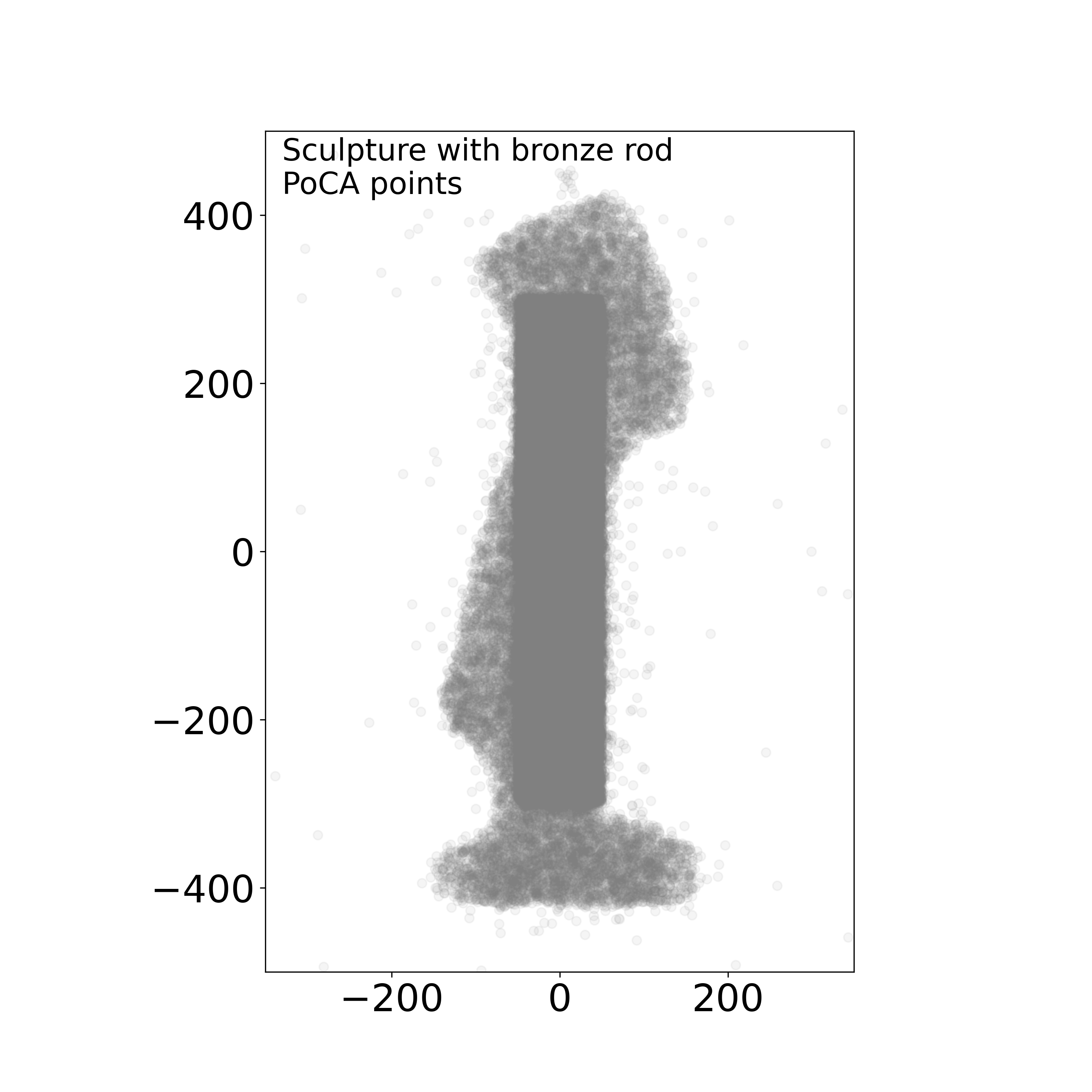}
    \caption{Distribution of POCA points, projected to a 2D plane for clarity, in three simulated scenarios: (left) actual wooden sculpture, (middle) with a cylindrical cavity, (right) with a cylindrical bronze rod.}
    \label{fig:poca-points}
\end{figure}

Scattering muography is based on the measurement of muon deflections when passing through an object. The deflection angle
is measured by extrapolating the incoming and outgoing trajectories observed by the two trackers and determining their point of closest approach (POCA). 
This approach relies on the interpretation of a POCA point as the actual place where the muon had a single high-energy elastic interaction with a nucleus, neglecting the occurrence of other electromagnetic interactions along its trajectory, which is rough approximation of reality but has proven to be effective in many applications.
Figure~\ref{fig:poca-points} shows the distribution of POCA points obtained in the three simulated scenarios denoted as I (a, b, c) in Table~\ref{tab-Simulation}. These plots are based on 5 million muons, roughly corresponding to an acquisition time of $\sim 8$ hours, and they show how challenging it is to find a cavity within this kind of statue, as opposed to finding a high-density insertion.




\begin{figure}[H]
\centering
\begin{minipage}{0.32\textwidth}
\includegraphics[width=\textwidth]{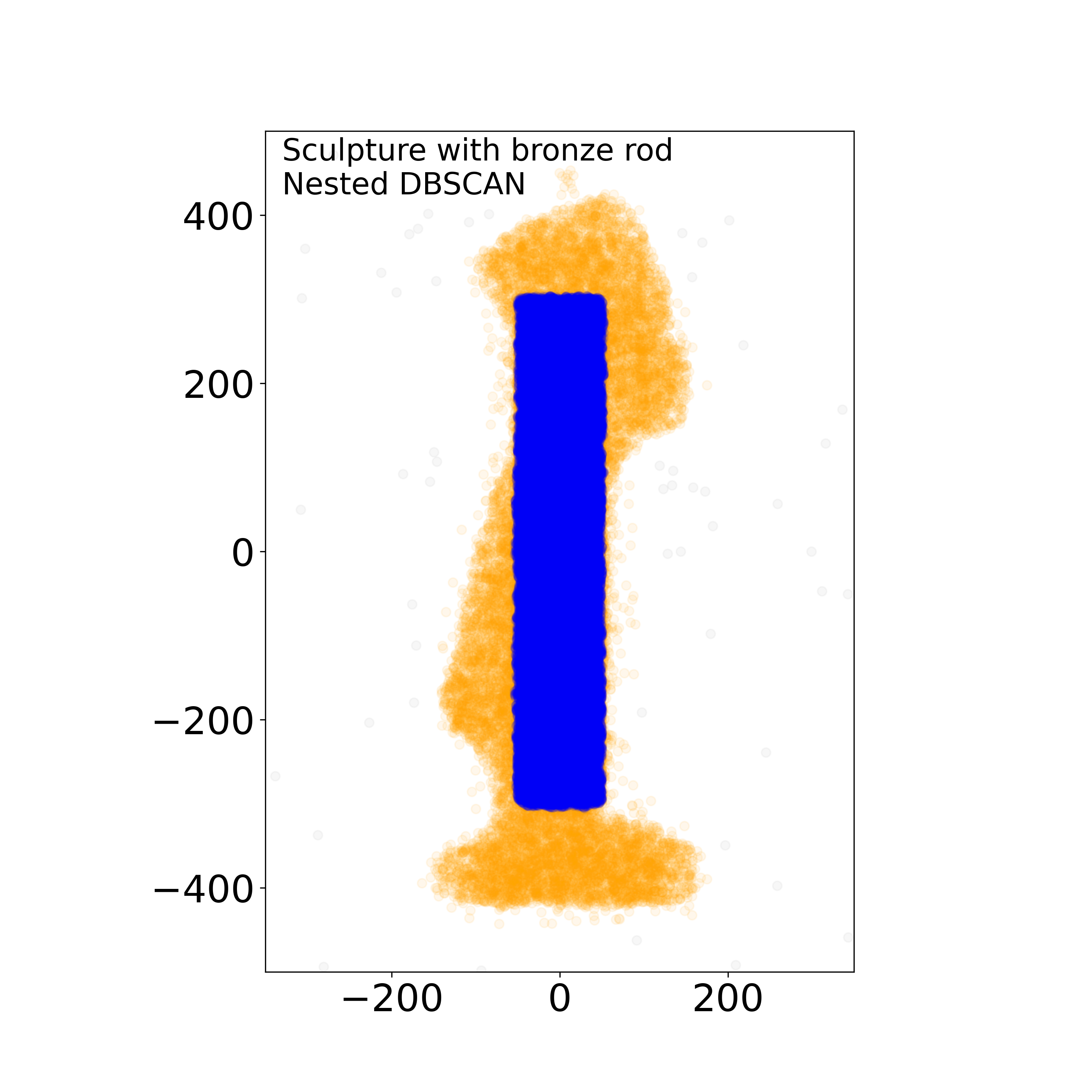}
\caption{Using DBScan clustering algorithm (blue: hidden copper bar, yellow: wood statue).}
\label{DBScan}
\end{minipage}\hspace{1pc}
\begin{minipage}{0.6\textwidth}
\includegraphics[width=0.49\textwidth]{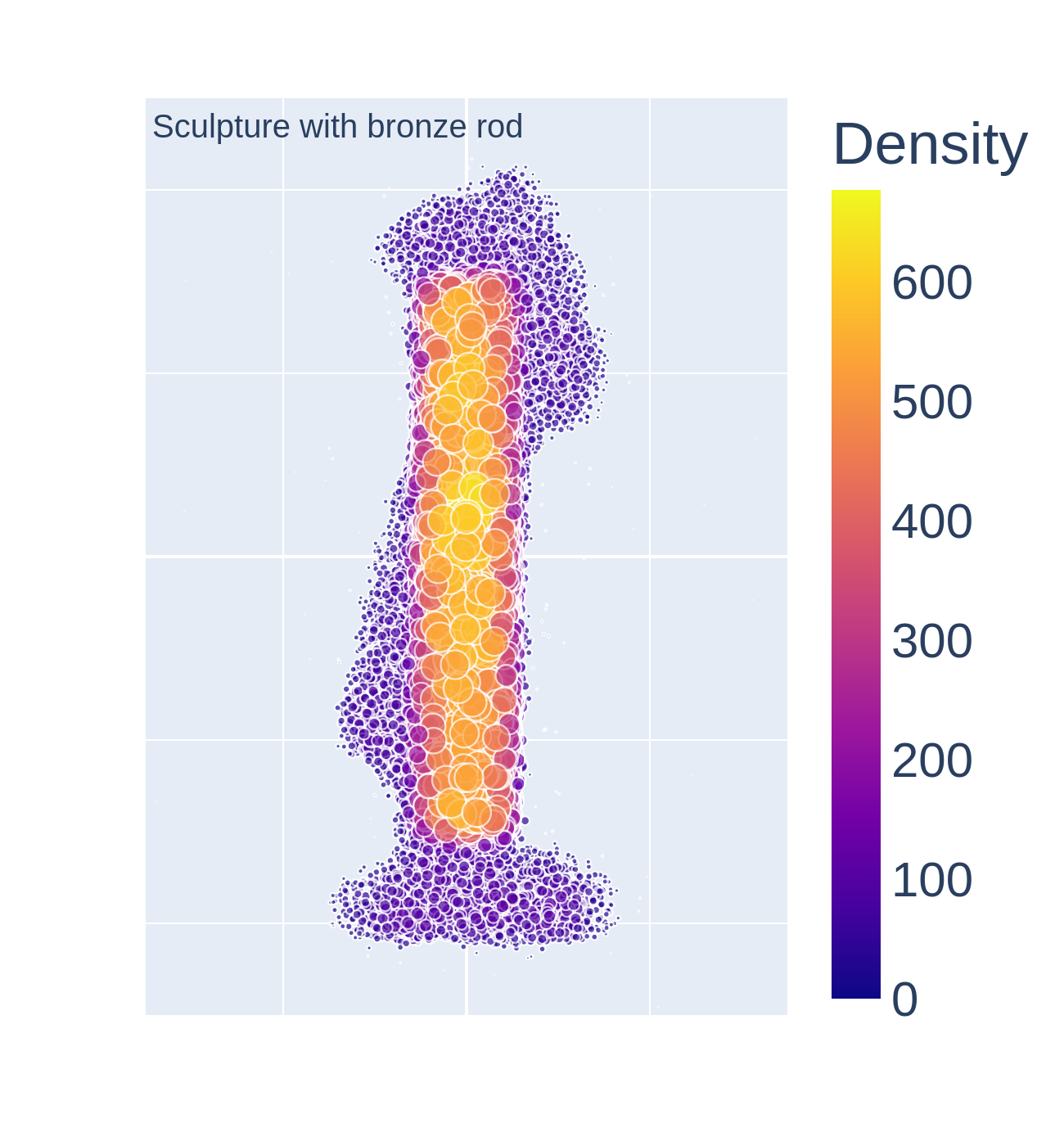}
\includegraphics[width=0.49\textwidth]{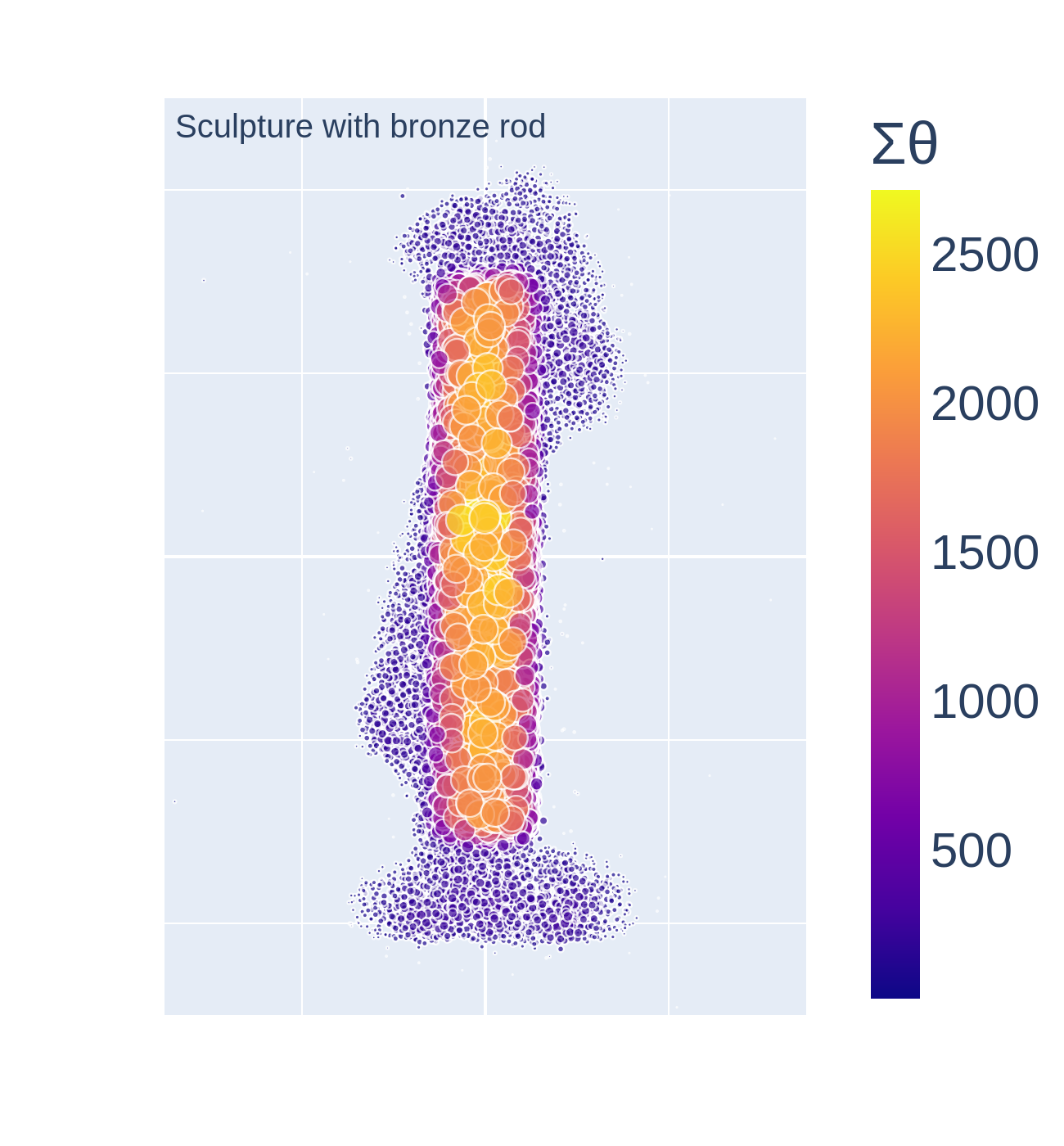}
\caption{Using Neighborhood Sum clustering algorithm (orange: hidden copper bar, magenta: wood statue), taking as input only the POCA positions (left) or the scattering angles (right).}
\label{Neighborhood}
\end{minipage}  
\end{figure}

The output of the muon-scattering reconstruction algorithm is a 3D distribution of POCA points, each associated to a scattering angle. Based on those raw data, some clustering algorithms can be used in order to discriminate between different material densities and elements.  
At present two methods are employed to analyze the object's content: DBSCAN~\cite{DBScan} and a Neighborhood Sum~\cite{kumarcomparative} algorithm. DBSCAN only relies on the density of the POCA points, while the Neighborhood Sum method can consider both the density of POCA points and the scattering angle of the tracks. We apply DBSCAN in two steps, to first remove noise points and then separate the volumes corresponding to different materials using tighter clustering criteria; the result is shown in Fig.~\ref{DBScan}. 
We applied the Neighborhood Sum without (Fig.~\ref{Neighborhood} left) and with (Fig.~\ref{Neighborhood} right) considering the additional information from the scattering angles, and we obtain in both cases a good discrimination of the two materials. This discrimination is not as precise as DBSCAN, however this method is more appropriate for scenarios with low exposure times, where POCA points are scarce, and the quantitative results it provides are still reliable.


{\bf Absorption reconstruction}

\begin{figure}[H]
\centering
\begin{minipage}{13pc}
\includegraphics[width=13pc]{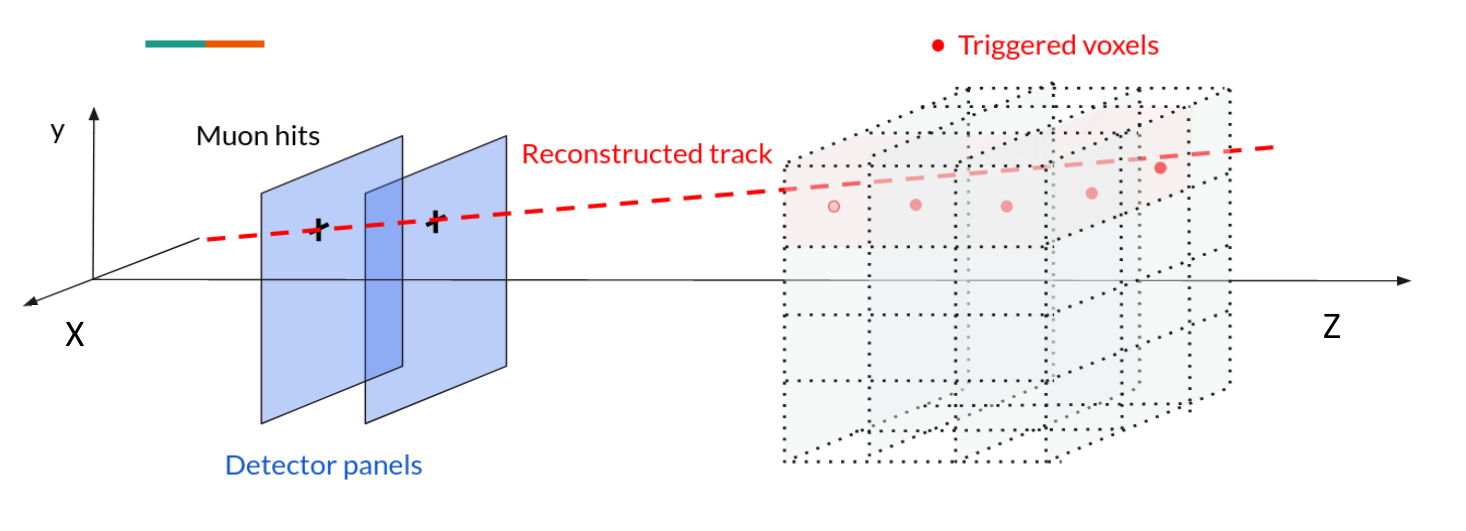}
\caption{Sketch for the back-projection algorithm.}
\label{absorption_rec}
\end{minipage}\hspace{1pc}
\begin{minipage}{13pc}
\includegraphics[width=12pc]{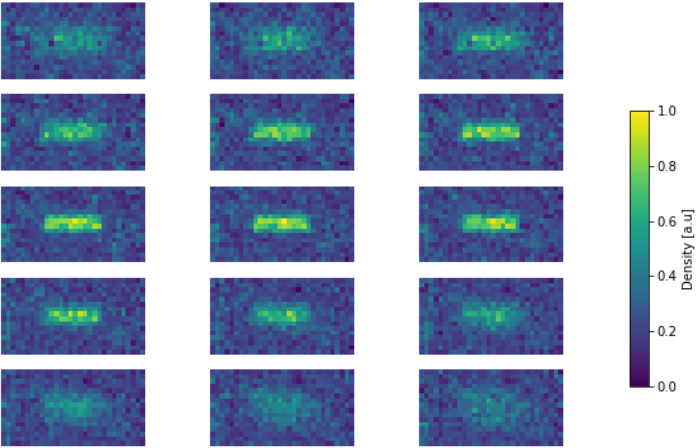}
\caption{Transmission map slice by slice.}
\label{transmassion}
\end{minipage}  
\end{figure}

With the scenario II described in Table~\ref{tab-Simulation}, we explore a challenging regime in which the statue is very big for scattering muography and very small for absorption muography.  
It is customary in this method, when the volume of interest is very distant from the detector (e.g. when imaging the summit of a volcano), to approximate the latter with a point, meaning that only the zenith and azimuth angles ($\theta,\phi$) are important while the entry point of the muon in the detector is not. 
However, to study human-sized sculptures we have in general the possibility to position the detectors very close to the statue, in order to maximize the resolution within the object, and this approximation is no longer valid. 
For this study we develop a custom back-projection reconstruction algorithm inspired by the methods of Refs.~\cite{ASR,BonechiBP}. As illustrated in Fig.\ref{absorption_rec}, we extrapolate each muon track onto a voxelized volume, and we count the number of times a voxel is hit by this backprojected trajectory. 
Figure~\ref{transmassion}, based on the equivalent of two hours of data acquisition, shows the 3D transmission map slice by slice in the voxelized volume after selecting only muons with $E<800$ MeV, assuming that the detector setup also contains a way to discriminate the muons above and below this energy threshold. Energy discrimination can be achieved cheaply by introducing a passive absorber before the last detector layer, used as a veto for energetic muons, or more precisely by combining absorption, scattering, time of flight, or other variables.

\section{Conclusion and prospects}
\label{sec:conclusion}

In this paper, we outlined the strengths and limitations of muography for cultural heritage applications, 
and detailed a preliminary simulation study with both scattering and absorption muography for the imaging of statues with different size and containing different hidden materials. 
The next step will be a systematic comparison of several more scenarios, in terms of material and size of the statue and of the hidden volume. We will also take into account various realistic scenarios for the detector resolution and the geometry of the setup (e.g. distances between planes), to identify the best trade-off between cost and statistical identification power, in preparation for actual measurements with test objects.

Muography is inexpensive and portable; thanks to the muon penetration power, it is complementary to other imaging methods. Absorption and scattering have complementary strengths and weaknesses, but some limitations are in common for both: long acquisition times are necessary, due to the relatively low natural rate, and muon direction and energy cannot be controlled.

At the workshop, we were invited to comment on what an artificial muon beam could do for this kind of studies. 
One could, indeed, overcome the aforementioned muography drawbacks by using a muon beam where both muon energy and direction could be controlled. Even a modest precision in these two variables and a modest beam luminosity, by accelerator standards, would allow to do much better than with muons from cosmic rays. 
The beam energy could be optimized a priori based on the size and on the main material of the object; if the inner composition is completely unknown, particularly interesting is the possibility to scan the same object with beams at various energies. 
We would benefit most if a transportable artificial muon source became accessible. 
However, this would bring radiological hazards, like all methods based on an artificial particle source, because of the byproducts of the collisions needed in order to produce muons and antimuons.

\section*{Acknowledgements}
We are indebted to Tim De Kock of the Antwerp Cultural Heritage Sciences (ARCHES) department at the University of Antwerp, Judy De Roy and Sam Huysmans of the Royal Institute for Cultural Heritage (KIK-IRPA), and Matthieu Boone of the Ghent University Centre for X-ray Tomography at the University of Ghent, for their guidance on the definition of the targets of interest for this potential application of muography.
We thank the Africa Museum of Tervuren and the project TOCOWO (\url{https://tocowo.ugent.be/}) for the model of a wooden statue (Figure~\ref{statue}). 
This work was partially supported by the Fonds de la Recherche Scientifique - FNRS under Grants No. T.0099.19 and J.0070.21, and by the EU Horizon 2020 Research and Innovation Programme under the Grant Agreements No. 822185 (``INTENSE'') and No. 101021812 (``SilentBorder'').


\bibliographystyle{IEEEtran}
\bibliography{main}

\end{document}